\documentstyle{pasj00}

\begin{document}

\SetRunningHead{A.Imada et al.}{The 2003 Superoutburst of an SU UMa-type
Dwarf Nova, GO Comae Berenices}

\title{The 2003 Superoutburst of an SU UMa-type Dwarf Nova, \\ GO Comae
Berenices}

\author{Akira \textsc{Imada}$^1$,Taichi \textsc{Kato}$^1$,Makoto
\textsc{Uemura}$^1$,Ryoko \textsc{Ishioka}$^1$,Thomas
\textsc{Krajci}$^2$ \\
Yasuo \textsc{Sano}$^3$,Tonny \textsc{Vanmunster}$^4$,Donn
\textsc{R.Starkey}$^{5}$,Lewis M.\textsc{Cook}$^6$
,Jochen \textsc{Pietz}$^7$ \\ 
Daisaku \textsc{Nogami}$^8$,Bill \textsc{Yeung}$^{9}$,Kazuhiro
Nakajima$^{10}$,Kenji \textsc{Tanabe}$^{11}$,Mitsuo \textsc{Koizumi}$^{11}$ \\
Hiroki \textsc{Taguchi}$^{11}$,Norimi \textsc{Yamada}$^{11}$,Yuichi
\textsc{Nishi}$^{11}$,Brian \textsc{Martin}$^{12}$,Ken'ichi
\textsc{Torii}$^{13}$ \\
Kenzo \textsc{Kinugasa}$^{14}$ and Christopher P.\textsc{Jones}$^{15}$}

\affil{$^1$Department of Astronomy,Faculty of Science, Kyoto University,
       Sakyo-ku, Kyoto 606-8502}
\email{a\_imada@kusastro.kyoto-u.ac.jp}
\affil{$^2$3933 Stockton Loop SE Albuquerque,New Mexico 87118-1104, USA}
\affil{$^3$VSOLJ,3-1-5 Nishi juni-jou minami,Nayoro,Hokkaido,Japan}
\affil{$^4$Center for Backyard Astrophysics(Belgium),Walhostraat
       1A,B-3401,Landen,Belgium}
\affil{$^{5}$AAVSO,2507 County Road 60,Auburn,Auburn,Indiana 46706,USA}
\affil{$^6$Center for Backyard Astrophysics(Concord),1730 Herix
       Ct.Concord, California 94518,USA}
\affil{$^7$Rostocker Str.62,50374 Erftstadt,Germany}
\affil{$^8$Hida Observatory,Kyoto University,Kamidakara,Gifu 506-1314}
\affil{$^{9}$PO Box 105 Benson, Arizona 85602, USA}
\affil{$^{10}$VSOLJ,124 Isatotyo Teradani,Kumano,Mie,Japan}
\affil{$^{11}$Department of Biosphere-Geosphere Systems, Faculty of
       Informatics,Okayama University of Science, \\Ridaicho 1-1,Okayama
       700-0005}
\affil{$^{12}$King's University College, Department of Physics, 9125
       50th Street, Edmonton, AB T5H 2M1, Canada}
\affil{$^{13}$Department of Earth and Space Science,
Graduate School of Science,1-1 Machikaneyama-cho,\\ Toyonaka, Osaka
       560-0043 Japan}
\affil{$^{14}$Gunma Astronomical Observatory, 6860-86 Nakayama Takayama-mura,
       Agatsuma-gun, Gunma 377-0702}
\affil{$^{15}$BAA, Variable Star Section, 29 Buller Rood, Laindon, Essex
       SS15 6BA, England}


\maketitle

\begin{abstract}

 We photometrically observed the 2003 June superoutburst of GO Gom. The
 mean superhump period was 0.063059(13) d. The resultant data
 revealed that (1) the obtained light curve contained a precursor, (2) a
 plateau stage of the object lasted 8 days, which is remarkably
 shorter than that of ordinary SU UMa-type dwarf novae, and (3) the
 amplitude of the superoutburst was less than 5 mag, which is
 unpredictably small when taking into account the fact that the
 supercycle of GO Com is about 2800 days. In order to explain these
 anomalies, a mass elimination process from the accretion disk, such as
 evaporation, may play an important role.    
\end{abstract}

\section{Introduction}

   Dwarf novae are a class of cataclysmic variables (CVs) which are
close binary systems consisting of a white dwarf and a late-type
main-sequence secondary transferring gaseous matter via the Roche
lobe overflow. The transferred gas forms an accretion disk around the
white dwarf. The accretion disk gives rise to
various kinds of instabilities and variabilities (for a review, see e.g.,
\cite{war95book}, \cite{hel01book}). The two major instabilities are
thermal and tidal instabilities, which are responsible for dwarf
nova-type outbursts and superhumps, respectively (for a review, see
\cite{osa96review}). SU UMa-type stars, subclass of dwarf novae, satisfy
both instabilities.

 It is widely recognized that SU UMa-type dwarf novae have a subgroup,
 called WZ Sge-type dwarf novae [originally proposed by
 \citet{bai79wzsge}; see also \citet{dow81wzsge}, \citet{odo91wzsge},
 \citet{kat01hvvir}]. WZ Sge-type dwarf novae are one of the most
 enigmatic systems all over the dwarf novae (see e.g.,
 \cite{osa95wzsge}, \cite{las95wzsge}, \cite{min98wzsge}). There are
 only five objects that have been firmly classified into WZ Sge-type
 dwarf novae, HV Vir, AL Com, EG Cnc, RZ Leo, and WZ Sge
 itself. Recent paper \citep{kat01hvvir} lists possible candidates for
 WZ Sge-type dwarf novae. The
 mass ratios of these systems have been suggested to be much smaller
 than those of SU UMa-type dwarf novae.\footnote{RZ Leo, however, is
 regarded as a system with a rather high mass ratio in comparison with
 the other WZ Sge-type dwarf novae.} Criteria for WZ 
 Sge-type dwarf novae are that (1) a supercycle, which
 is a period between two succeeding superoutbursts, is over 5 years,
 much longer than those of SU UMa-type dwarf novae, (2) their outbursts
 show very large amplitudes exceeding 6 mag, and (3) near the maximum of a
 superoutburst, early superhumps, having doubly-peaked shapes and
 periods close to the $P_{\rm orb}$, are observed for about a few days
 before emergence of ordinary superhumps.\footnote{The early
 superhump is sometimes called orbital hump \citep{pat02wzsge} or early
 hump \citep{osa02wzsgehump}. The difference between these authors is
 originated from their interpretation of physical processes near the bright
 maximum} Historically, \citet{vog82atlas} pointed out that GO Com is a
 candidate for WZ Sge-type dwarf novae.\footnote{Note that criteria for
 TOADs in \citet{how95TOAD} are different from those for WZ Sge-type
 dwarf novae} However, the lack of observations has prevented us from
 clarification.
     
 GO Com was discovered by \citet{kow77gocomiauc} as an eruptive object on
 a Palomar plate taken on 1973 July 1.213. This variable was confirmed
 to be identical with the suspected variable star CSVS 1959 = SVS 382
 discovered by \citet{bel33gocom}. \citet{how90faintCV3} derived the
 orbital period of GO Com to be 95 min. 

   Being located extremely close to the north galactic pole and in the SA
   57 region, this object was independently recorded as a very blue
   object of $B$=18.1 (US 31) during a systematic survey of this region
   \citep{ush81gocom}. According to \citet{ush81gocom}, this object has
   one of the most unusual colors (U$-$B = $-$1.5) in this region. This
   object is also cataloged as a quasar candidate OMHR
   J125637.12+263644.2 \citep{mor95quasar}. 

   The object has been visually monitored by the members of the Variable
   Star Observers League in Japan (VSOLJ) since 1986, and only one short
   outburst was observed on 1989 May 30 (m$_v$=13.2) until 1995. This
   1989 outburst showed a rapid decline by about 1 mag d$^{-1}$,
   suggesting that the outburst would be a normal outburst of a
   short-period dwarf nova. Although there were unavoidable seasonal
   gaps in monitoring, the extremely low outburst frequency of this
   dwarf nova has been established by this observation. The apparent
   absence of long outbursts, infrequent short outbursts, and possible
   excursion to a low state (as indirectly inferred from the literature)
   have suggested similarity to the unusual short-period dwarf nova BZ UMa
   (\cite{wen82bzuma}; \cite{kal86bzumaiauc};
   \cite{rin90bzuma};\cite{jur94bzuma}; \cite{kat99bzuma};
   \cite{neu02bzumaproc}). The relatively strong X-ray emission detected
   by ROSAT \citep{ver97ROSAT} also suggests the unusual nature of GO Com.

   In 1990, \citet{kat90gocom} conducted first-ever time-resolved CCD
   photometry of GO Com near its quiescence, and reported the possible
   presence of a periodicity of 33 min. \citet{how90faintCV3} found a
   possible 95-min periodicity by CCD photometry. Although this period was
   tentatively related to the possible orbital period, the short
   observation was not sufficient to conclude on the reality of this
   periodicity.  We have had to wait for more than a decade before the
   true periodicity is confirmed by the present observation. 

 The 1995 July outburst was detected by Vanmunster on 1995 July 16,
 which was confirmed by CCD observation on July 25 at $V$=16.6
 \citep{kat95gocom}. The most unusual feature of this outburst was that
 this outburst preceded a long outburst by 14 d. This `second'
 outburst very much looked like a superoutburst of SU UMa-type dwarf
 novae\citep{kat95gocom}, although the limited visibility hindered
 secure detection of superhumps. The object even underwent a
 rebrightening following this outburst. All of these observations
 suggested that GO Com is a genuine SU UMa-type dwarf novae, and that the
 1995 July outburst can be best interpreted as a precursor outburst to
 the following presumable superoutburst (cf. \cite{mar79superoutburst};
 \cite{kat97tleo}). 

   Since then, there have been only four outbursts on the VSNET\footnote{
$\langle$http://www.kusastro.kyoto-u.ac.jp/vsnet/$\rangle$} record:
1996 March, 1997 February, 1998 April, and 2003 June.  The outbursts
other than the last one were short and faint outbursts.

\section{Observation}
 The CCD observations were performed by the international VSNET \citep{VSNET}
 collaboration team at 14 sites. A summary of the observations and
 the equipment used in each site is listed in table 1.

 The images were dark-subtracted, flat-fielded and analyzed with the
aperture photometry. The Kyoto and RIKEN images were analysed using a
Java-based aperture and PSF photometry package developed by one of the
authors (TK). The Hida images were analysed using IRAF. The other images
were analyzed using AIP4WIN.

The magnitude scales of each observatory were adjusted to that of the
most abundant Tashkent system (TKr in table 1). The magnitudes of the
object were determined relatively using a local standard star,
GSC1995.488, whose constancy during the run was confirmed using a check
star, GSC1995.1151.

Heliocentric corrections to the observation times were applied before
the following analysis.

\section{Results}

\subsection{Long-term Monitoring of GO Com}

As mentioned above, extensive monitoring of the object has been
performed by many amateur astronomers. Data points which include
negative observations amount to more than 4000 since 1995. However,
despite the secure observations, only 2 superoutbursts have been
detected. Table 2 represents the records of outbursts of GO Com. One can
roughly estimate a supercycle of GO Com about 2800 days, judging from
the two succeeding superoutbursts in 1995 July and 2003 June.

\subsection{Light Curve}

The resultant light curve of the superoutburst is shown in figure 1. We
derived the decline rate of 0.15 mag d$^-$$^1$ during the plateau stage,
which lasted for 8 days, from HJD2452796 to HJD2452804. The most
significant fact is that the duration of the plateau stage is
considerably shorter than that of normal SU UMa-type dwarf novae, whose
duration is typically about 2 weeks (Table 3.7. in
\citet{war95book}).

After the termination of the steep decline, the object entered the
post-superoutburst stage during which the object stayed more than about
1 mag brighter than in quiescence ( B=18.1, mentioned above, and
typically B-V $\sim$ 0 in dwarf novae). A rebrightening feature is sometimes
observed during this stage, especially among SU UMa-type dwarf novae
with shorter orbital periods and WZ Sge-type dwarf novae. There is a
hint of a rebrightening feature around HJD2452811, however, the lack of
later data has hindered us from confirmation of rebrightening.  

\subsection{Superhumps}

The result of a period analysis is depicted in figure 2, applying the
 PDM method \citep{ste78pdm} to a data set covering the period from
 HJD2452795 to HJD2452801 after subtracting the linear decline trend. We
 obtained the mean superhump period $P_{\rm SH}$ of 0.06297(3) d.
 Figure 3 represents the phase-averaged mean superhump profile. The
 rapid-rise and slow-decline trend is characteristic of superhumps in SU
 UMa-type dwarf novae. This superhump period seems to be shorter than
 the orbital period derived by \citet{how90faintCV3}. Precise
 measurement of the orbital period of GO Com is urged.  
 
 In order to investigate variations of superhump profile, we also derived
 the daily-averaged profiles of superhumps, shown in figure 4. The
 superhump amplitude on HJD2452795, corresponding to the very day
 when the superoutburst started, was as large as 0.4 mag, which
 gradually decayed during the following 8 days. However, at the end of a
 plateau stage, superhumps clearly regrew. Similar phenomena were
 observed SU UMa-type dwarf novae with longer supercycle (see
 e.g.,\cite{bab00v1028cyg}).

\subsection{Superhump Period Change}

 We derived the timings of the superhump maxima mainly by eye. The
 accuracy of the eye estimates of the timings is around 0.001 d. The
 results are listed in table 3. A linear regression to the times yielded
 the following equation,

\begin{equation}
 HJD(max) = 2452795.1515(15) + 0.063059(13)\times E   
\end{equation}

 where $E$ is the cycle count since HJD2452795.1571.      
 By fitting the deviations of the observed timings from the calculation,
 we obtained the following equation,

\begin{equation}
 O - C = 0.0079(15) - 0.0005(1)\times E + 5.79(88)\times 10^{-6}\times E^2
\end{equation}

 The quadratic term corresponds to $\dot{P}$/$P$ =
 1.8$\times$10$^{-4}$ (see figure 5). This means that the superhump
 period became longer, as shown in some SU UMa-type dwarf novae with
 shorter orbital periods \citep{kat01hvvir}.

 Figure 5 also gives us a hint of a late superhump, around HJD
 2452802. However, it is difficult to examine it because of lacking of
 data.
  
\subsection{Precursor Outburst}

 SU UMa-type dwarf novae sometimes undergo a superoutburst
 following a normal outburst (\cite{mar79superoutburst}). We call such a
 normal outburst precursor. Thanks to the excellent coverage, the
 early stage of the present outburst is considered as a precursor before the
 superoutburst. Figure 1 indicates a decline of the precursor on
 HJD2452794.
 
 Our calculation revealed that the mean decline rate of magnitude during
 the precursor was 0.74 mag d$^-$$^1$. Figure 6 represents the precursor
 profile folded by P=0.063059 d after subtracting the decline
 trend. Judging from figure 6, there exists no superhump
 profile as is seen in a superoutburst of SU UMa-type dwarf novae.

\section{Discussion}

\subsection{The nature of GO Com}

As mentioned above, during a superoutburst, the amplitude and plateau stage
of ordinary SU UMa-type dwarf novae are 4$\sim$6 mag and $\sim$2 weeks,
respectively \citep{war95book}. On the other hand, during the 2003
superoutburst of GO Com, the amplitude and plateau stage of the object
are $\sim$5 mag and $\sim$8 days, respectively. Especially, the plateau
stage of GO Com is remarkably shorter than that of typical SU UMa-type
dwarf novae. This may indicate that the accumulated mass on the
accretion disk in quiescence is low in GO Com.

According to the disk instability theory, if an outburst is triggered
in an accretion disk with a relatively low mass, the system is expected
to have a short supercycle. However, in the case of GO Com, the
observations suggest a long supercycle of $\sim$2800 days and a
moderately short superhump period of 0.063059 days. These results are
beyond the framework of the standard disk instability model.

In order to account for the observations, we propose that high
efficiency of mass accretion onto the white dwarf is realized during
quiescence, This scenario was originally suggested by
\citet{las95wzsge}: the short plateau stage and the long supercycle are
due to inner truncation of the accretion disk. Similar interpretation is
discussed on LL And by \citet{kat04lland}. LL And is an SU UMa-type
dwarf nova with an extremely long supercycle ($\sim$5000 d), comparable
to that of WZ Sge-type dwarf novae. The obtained light curve of the 1993
superoutburst of LL And is remarkably similar to that of GO Com in terms
of a small superoutburst amplitude and a short plateau stage. Table 4
demonstrates fundamental characteristics of GO Com and LL And. In
conjunction with our observations, we conclude that the inner part of
the accretion disk may be truncated, as modeled by \citet{las95wzsge}.

Note, however, that various durations of superoutbursts have been observed
in certain objects(e.g. SW UMa,\citet{how95swumabcumatvcrv}). Further observations
are needed in order to clarify superoutburst propeties of GO Com.

\subsection{Superhump period change}

In general, the period of superhump in SU UMa-type dwarf novae
decreases during a superoutburst
(e.g.\cite{war85suuma}; \cite{pat93vyaqr}). The obtained $\dot{P}$/$P$
has a rather common negative value
($\sim$-5$\times$10$^{-5}$)\citep{kat03v877arakktelpucma}, which has
been generally ascribed to decreasing apsidal motion due to a decreasing disk
radius. \citet{osa85SHexcess} pointed out mass depletion from the disk
results in the decrease of superhump period.

However, recent observations reveal that the minority of SU UMa-type
dwarf novae has positive $\dot{P}$/$P$ (\cite{sem97swuma};
\cite{nog98swuma}; \cite{bab00v1028cyg}; \cite{kat01wxcet};
\cite{kat02v592cas}; \cite{ole03v1141aql}; \cite{ole03ksuma};
\cite{nog04vwcrb}).\footnote{All WZ Sge-type dwarf novae show positive
$\dot{P}$.} \citet{kat01hvvir} suggested that an increase in superhump
periods may be related to a low mass ratio and/or a low mass transfer
rate. EI Psc \citep{uem02j2329}, however, obviously violates these
regime: \citet{uem02j2329letter} derived a high mass ratio (q =
$M_{\rm 2}$/$M_{\rm 1}$ = 0.19${\pm}$0.02) by using the empirical
relation between superhump excess and mass ratio
\citep{pat01SH}. Spectroscopic observations for EI Psc also indicate the
high mass transfer rate (\cite{hu98j2329}; \cite{wei01j2329}). It is
suggested that another condition may be required for explaining positive
$\dot{P}$.

Figure 7 demonstrates a $\dot{P}$/$P$ diagram of SU UMa-type dwarf
novae (including WZ Sge-type dwarf novae). Note that the obtained
$\dot{P}$/$P$ of GO Com is the second largest ever known. In addition, the
location of GO Com in figure 7 are far from the general trend from
upper-left to lower-right, implying a peculiar nature of GO Com.

\subsection{Precursor and before the maximum}

Observations before the maximum allow us to test the validity of a
refinement of the thermal-tidal instability model developed by
\citet{osa03DNoutburst}. \citet{osa03DNoutburst} suggested that a
superoutburst which is accompanied with a precursor should develop
superhumps after the termination of the precursor, i.e., as soon as a
superoutburst is triggered, superhumps emerge. In this point, our
observations are in agreement with the prediction by
\citet{osa03DNoutburst}.

We also investigated the possibilities of the presence of superhumps
during the precursor. \citet{kat97tleo}
found a hump-like profile with an amplitude smaller than 0.1 mag during
a precursor of the 1993 superoutburst in T Leo. He suggested that this
modulation and the superhump were caused by the same origin, based on
the fact that the ephemeris of the superhump-maximum timings during the
superoutburst was applicable to that of the modulations during the
precursor. This implies that a tidal instability developed before the
main superoutburst. However, in the case of GO Com, there is no evidence
of superhumps during the precursor. The reason for the difference is
left as an open question.

\subsection{Relation to WZ Sge-type dwarf novae}

As mentioned above, historical records indicate that the supercycle of
GO Com is about 2800 day, which is considerably longer than well-known
SU UMa-type dwarf novae (200-350 d) (\cite{nog97sxlmi},
\cite{kat03hodel}). Thus, from the point of supercycle, GO Com is
similar to WZ Sge-type dwarf novae, although their supercycles are much
longer than that of GO Com.

However, GO Com underwent normal outbursts more frequently than
superoutbursts. Observations reveal that WZ Sge-type dwarf novae enter a
normal outburst very infrequently.\footnote{There is no record of a
normal outburst for WZ Sge itself.} Theories predict that
an extremely low viscosity on an accretion disk during quiescence leads
WZ Sge-type dwarf novae to have a long dormancy (\cite{osa95wzsge}). In
the case of GO Com, assuming the analogy with WZ Sge-type dwarf
novae, rather frequent normal outbursts may not be explained. Thus, the
viscosity in the accretion disk is not necessarily low during quiescence.

We also investigated the possibility of the existence of an early superhump
on HJD 2452795, one day before the maximum. However, the profiles
we detected was not doubly peaked, but singly peaked. Therefore, the
modulations on HJD 2452795 were definitely ordinary superhumps. In
addition, precursor outbursts are not shown among WZ Sge-type dwarf
novae. That's why we disqualify GO Com as WZ Sge-type dwarf novae.

\section{Summary}

 Our photometrical study in this paper is summarized as follows.

1. GO Com was proved to be an SU UMa-type dwarf nova with superhumps
  whose period is specified as 0.063059 d.

2. The plateau stage of the superoutburst continued only 8 days, during
  which the period of superhump became longer.

3. A precursor was followed by the superoutburst. Our detailed
  analysis finds no evidence for superhumps during the precursor, which
  is in contrast with the results of \citet{kat97tleo} about T Leo.

4. Judging from long-term monitoring, the supercycle of GO Com is about
  2800 d, whereas the amplitude of the 2003 superoutburst was less than
  5 mag. In conjunction with its plateau stage of GO Com (8 days), the
  total mass accretion onto the white dwarf might be low. 

\vskip 3mm

We are grateful to many VSNET observers who have reported vital observations.
We are particularly grateful to Christopher Jones for promptly reporting
the outbursts through VSNET. This work is supported by the Grant-in-Aid
for the 21st Century COE "Center for Diversity and Universality in
Physics" from the Ministry of Education, Culture, Sports, Science and
Technology (MEXT) of Japan. This work is partly supported by a grant-in
aid [13640239 (TK), 14740131 (HY), 16740121 (KK)] from the Ministry of
Education, Culture, Sports, Science and Technology. Part of this work is
supported by a Research Fellowship of the Japan Society for the Promotion of
Science for Young Scientists (MU, RI). This research has made use of the
Digitized Sky Survey produced by STScI, the ESO Skycat tool, and the
VizieR catalogue access tool. 

\begin{figure*}
\begin{center}
\resizebox{160mm}{!}{\includegraphics{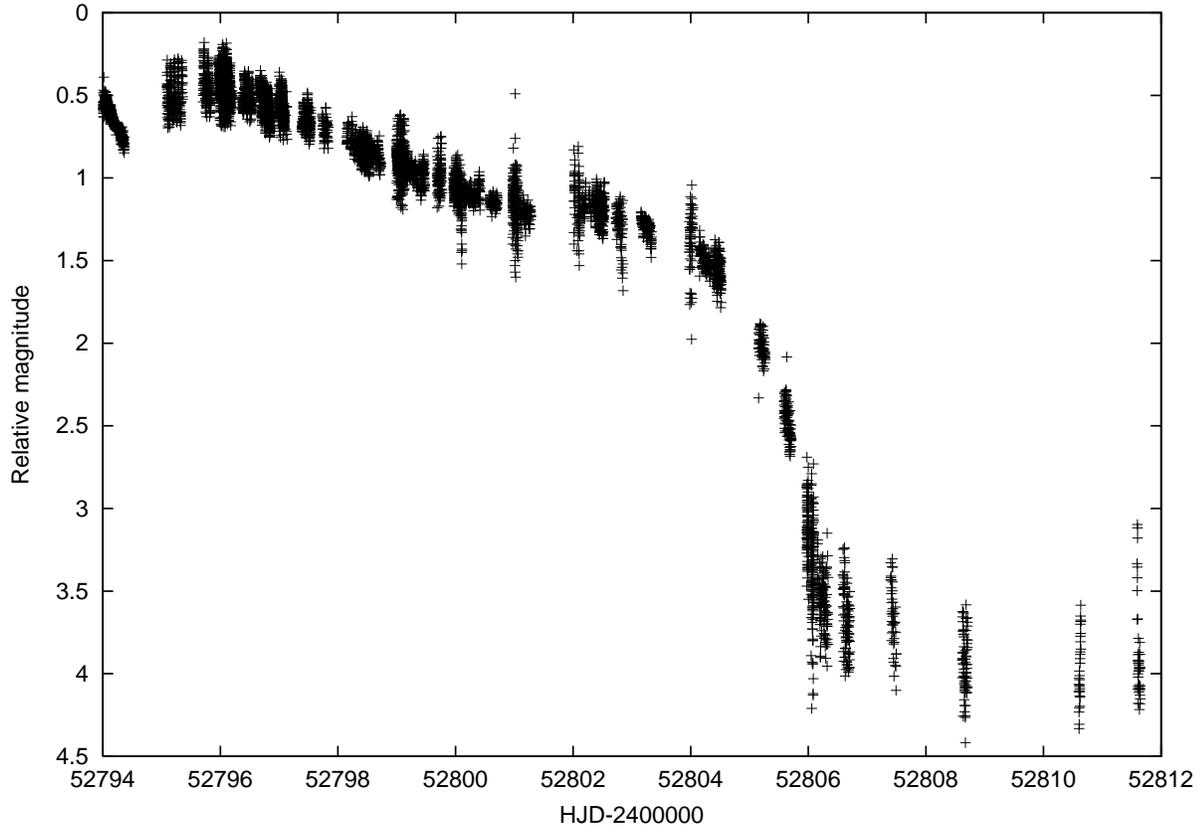}}
\caption{The whole light curves of the 2003 superoutburst of GO Com. The
 zero point corresponds to magnitude 12.80, on a system close to R. On
 HJD 2452794, there shows a precursor with a decline rate of 0.74 mag
 d$^{-1}$.}
\label{}
\end{center}
\end{figure*}

\begin{figure*}
\begin{center}
\resizebox{80mm}{!}{\includegraphics{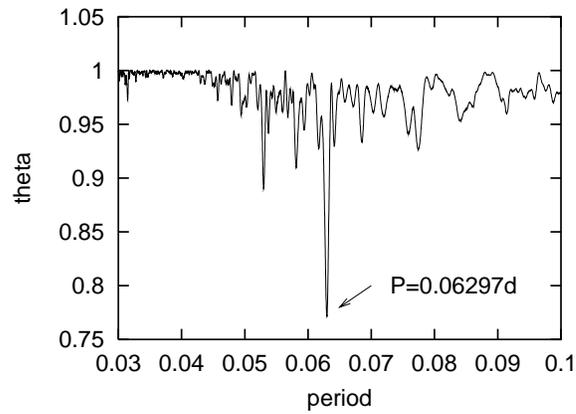}}
\caption{Theta diagram of a PDM period analysis of the data obtained
 during the plateau stage. The minimum point indicates 0.06297 d as the
 best-estimated superhump period.}
\end{center}
\end{figure*}

\begin{figure*}
\begin{center}
\resizebox{80mm}{!}{\includegraphics{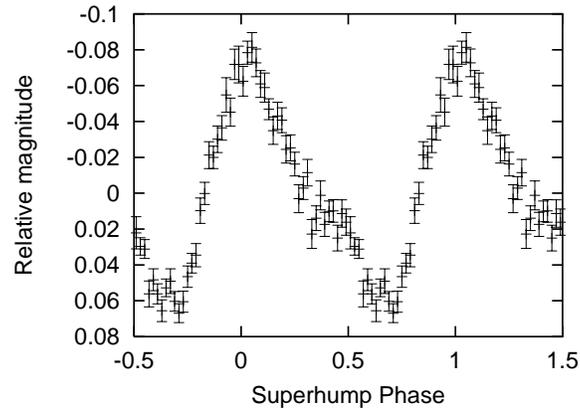}}
\caption{Phase-averaged light curve of GO Com superhumps, covering
 between HJD2452795 and HJD2452801, folded by 0.06359 d. The vertical
 and the horizontal axis denote the relative magnitude and the phase,
 respectively. A rapid rise and slower decline, which is a typical
 feature of superhumps, are shown. }
\end{center}
\end{figure*}

\begin{figure*}
\begin{center}
\resizebox{42mm}{!}{\includegraphics{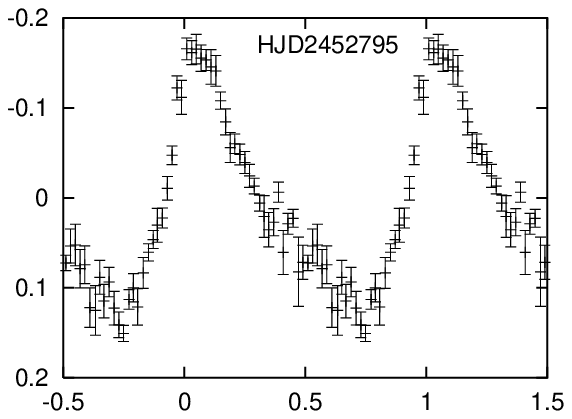}}
\resizebox{42mm}{!}{\includegraphics{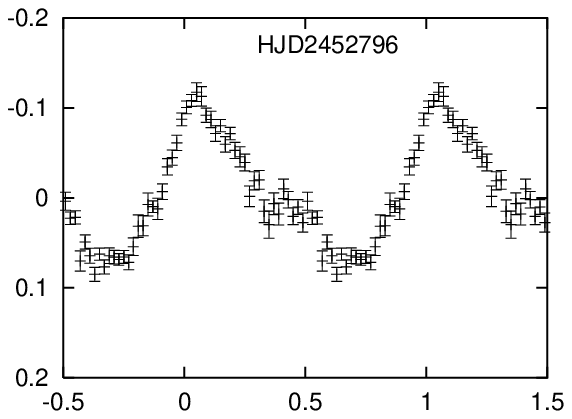}}
\resizebox{42mm}{!}{\includegraphics{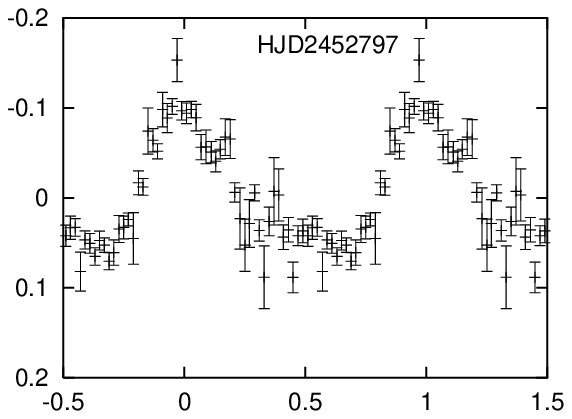}}
\resizebox{42mm}{!}{\includegraphics{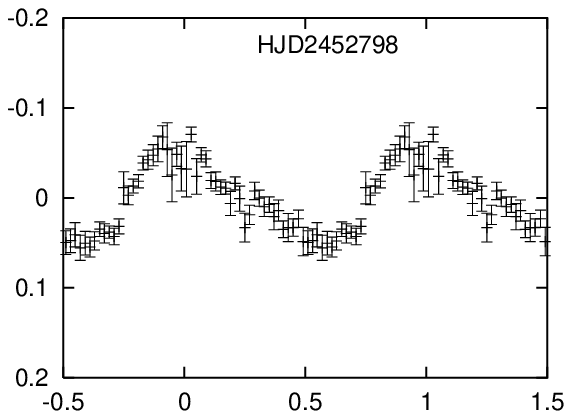}}
\resizebox{42mm}{!}{\includegraphics{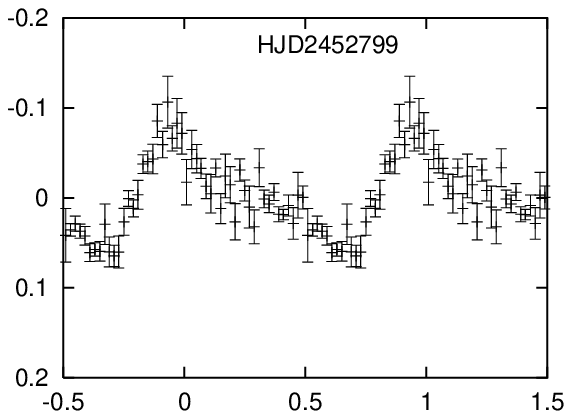}}
\resizebox{42mm}{!}{\includegraphics{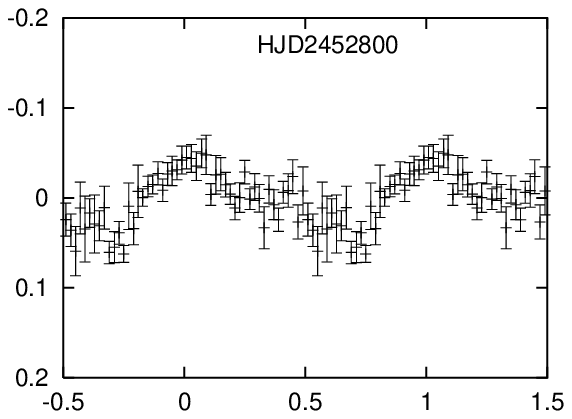}}
\resizebox{42mm}{!}{\includegraphics{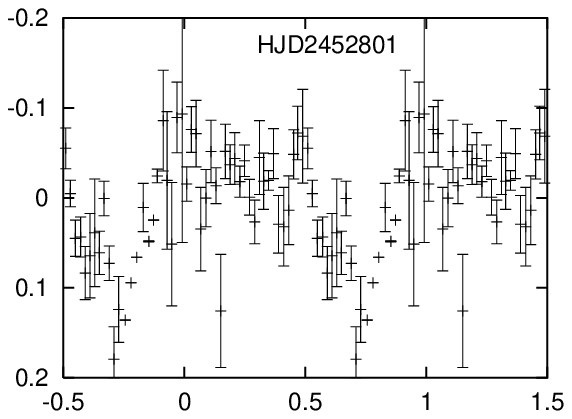}}
\resizebox{42mm}{!}{\includegraphics{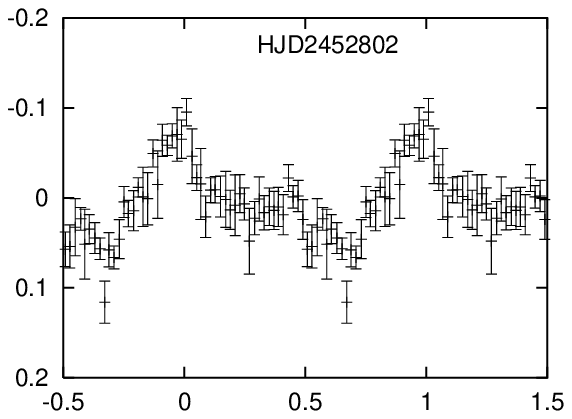}}
\resizebox{42mm}{!}{\includegraphics{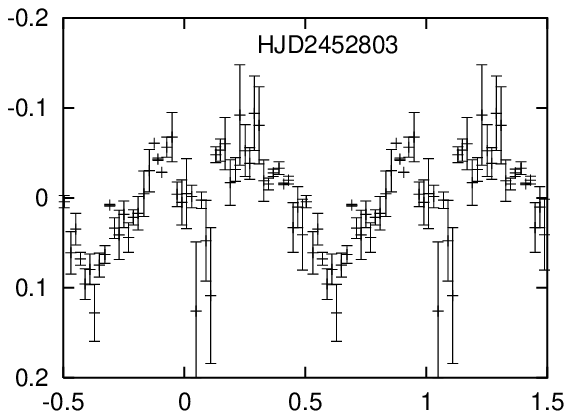}}
\resizebox{42mm}{!}{\includegraphics{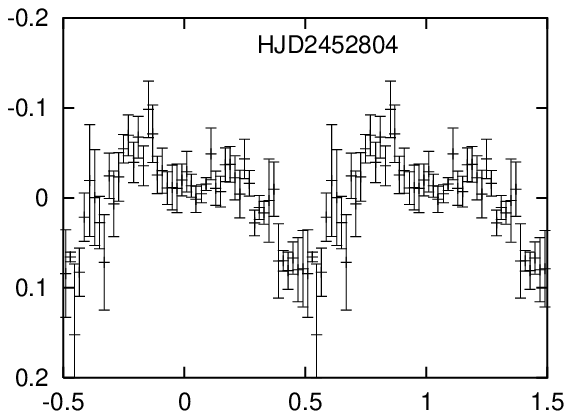}}
\resizebox{42mm}{!}{\includegraphics{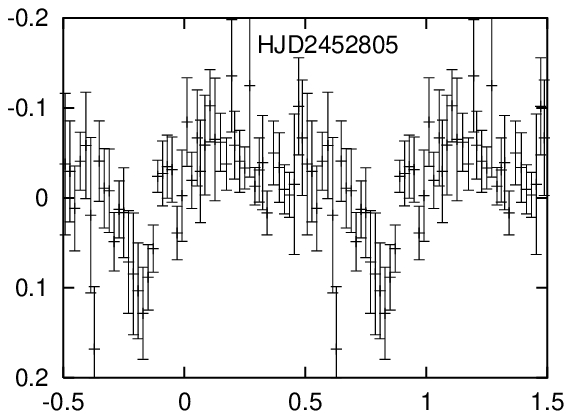}}
\resizebox{42mm}{!}{\includegraphics{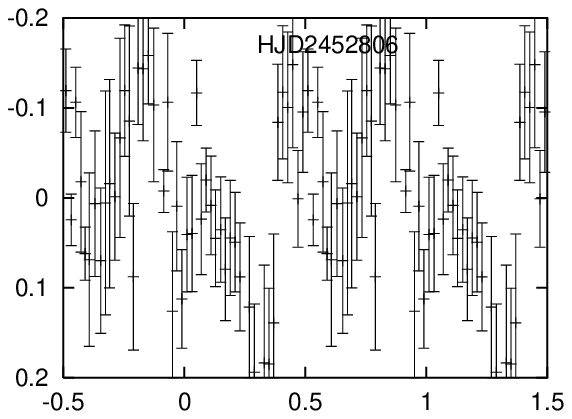}}
\caption{Phase-averaged light curves of GO Com superhumps. The abscissa
 and the ordinate are the phase of the superhump period and the
 differential magnitude, respectively. }
\label{}
\end{center}
\end{figure*}

\begin{figure*}
\begin{center}
\resizebox{80mm}{!}{\includegraphics{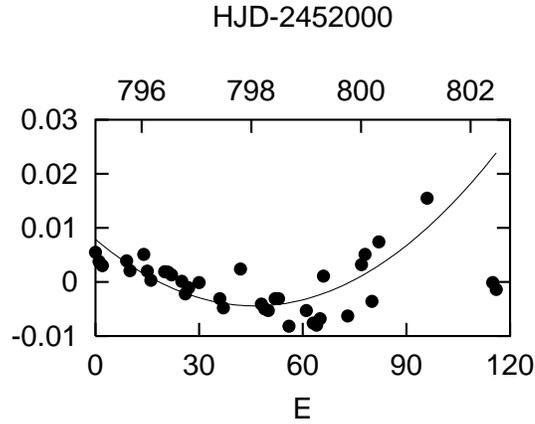}}
\caption{O-C diagram of superhump maxima. The abscissa and the ordinate
 denote the cycle count and $O-C$ (d) using equation (1),
 respectively. Typical error is 0.001 in each point. The quadratic fit
 corresponds to equation(2). There is a hint of a late superhump around
 HJD 2452802.}
\label{}
\end{center}
\end{figure*}

\begin{figure*}
\begin{center}
\resizebox{80mm}{!}{\includegraphics{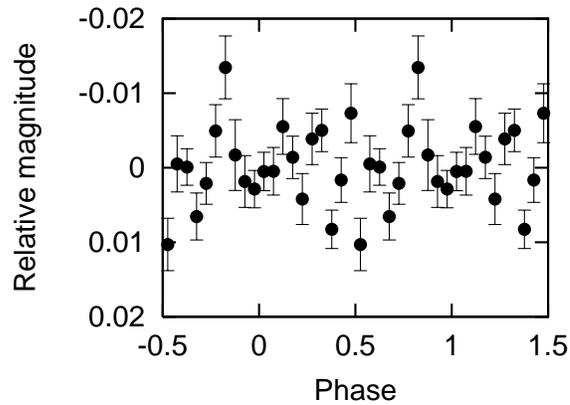}}
\caption{Phase-averaged superhump profile on HJD2452794, corresponding
 to the precursor phase. There is no feature of superhump. This result
 was in contrast with the 1993 superoutburst of T Leo \citep{kat97tleo}
 in which hump-like modulations, corresponding to the superhump period,
 was observed even in precursor. }
\label{}
\end{center}
\end{figure*}

\begin{figure*}
\begin{center}
\resizebox{80mm}{!}{\includegraphics{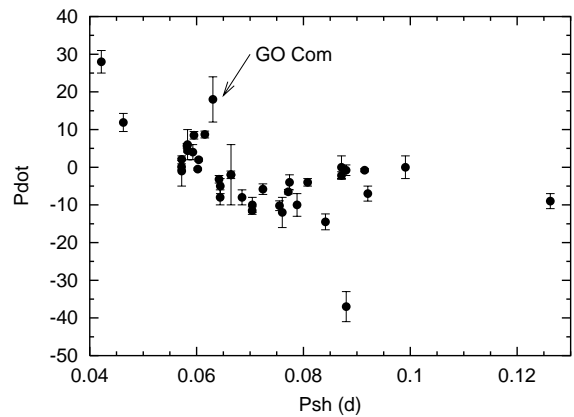}}
\caption{$P_{\rm dot}$=$\dot{P}_{\rm SH}$/$P_{\rm SH}$ versus $P_{\rm SH}$ diagram. The data of GO Com are added to the original figure (figure 8 in \cite{ish03hvvir}).} 
\label{}
\end{center}
\end{figure*}

\begin{longtable}{cccccc}
\caption{Log of observations}

\hline\hline
Date & HJD(start)$^*$ & HJD(end) & N$^\dagger$ & Exp(s)$^\ddagger$ &
 Observer \\
\hline
2003 June 3 & 52793.9623 & 52794.1228 & 170 & 60 & TO  \\
 & 52794.0083 & 52794.0213 & 27 & 30 & GU \\
 & 52794.0407 & 52794.1488 & 461 & 10 & ND \\
 & 52794.1546 & 52794.3686 & 254 & 60 & TKr \\
2003 June 4 & 52795.0940 & 52795.1571 & 79 & 30 & KYO \\
 &52795.1515 & 52795.3608 & 141 & 120 & TKr \\
2003 June 5 & 52795.6501 & 52795.7196 & 78 & 30 & BY \\
 & 52795.7068 & 52795.8395 & 213 & 30 & LC \\
 & 52795.9731 & 52796.0364 & 233 & 30 & NH \\
 & 52795.9916 & 52796.1208 & 168 & 30 & KYO \\
 & 52796.0051 & 52796.1106 & 467 & 10 & ND \\
 & 52796.0360 & 52796.1177 & 85 & 60 & TA \\
 & 52796.0408 & 52796.0843 & 499 & 3 & SY \\
 & 52796.1519 & 52796.2097 & 25 & 120 & TKr \\
 & 52796.3758 & 52796.5475 & 270 & 40 & TV \\
2003 June 6 & 52796.6002 & 52796.7288 & 94 & 90 & DS \\
 & 52796.6675 & 52796.8522 & 513 & 30 & BY \\
 & 52796.7079 & 52796.8725 & 262 & 30 & LC \\
 & 52796.9587 & 52797.0507 & 100 & 30 & NH \\
 & 52796.9613 & 52797.1570 & 87 & 60 & TO \\
 & 52797.0199 & 52797.0878 & 408 & 3 & SY \\
 & 52797.1044 & 52797.2231 & 24 & 30 & KYO \\
 & 52797.3882 & 52797.5218 & 150 & 30 & JP \\
 & 52797.4377 & 52797.5400 & 120 & 60 & TV \\
2003 June 7 & 52797.7359 & 52797.8367 & 57 & 120 & MA \\
 & 52798.1500 & 52798.3598 & 115 & 120 & TKr \\
 & 52798.3781 & 52798.5449 & 273 & 40 & TV \\
 & 52798.3858 & 52798.5346 & 173 & 30 & JP \\
2003 June 8 & 52798.6178 & 52798.7296 & 85 & 90 & DS \\
 & 52798.6947 & 52798.7165 & 15 & 30 & LC \\
 & 52798.9797 & 52799.1420 & 189 & 30 & KYO \\
 & 52798.9825 & 52799.1028 & 442 & 10 & ND \\
 & 52798.9985 & 52799.1212 & 434 & 20 & SY \\
 & 52799.0367 & 52799.1069 & 80 & 60 & TA \\
 & 52799.1528 & 52799.3560 & 139 & 60 & TKr \\
 & 52799.4082 & 52799.4738 & 102 & 40 & TV \\
2003 June 9 & 52799.6882 & 52799.7583 & 157 & 30 & LC \\
 & 52799.9696 & 52800.0387 & 419 & 20 & SY \\
 & 52800.1506 & 52800.3421 & 132 & 60 & TKr \\
 & 52800.3789 & 52800.4222 & 40 & 45 & JP \\
2003 June 10 & 52800.5842 & 52800.7114 & 104 & 90 & DS \\
 & 52800.9717 & 52801.0623 & 273 & 20 & SY \\
 & 52801.1504 & 52801.2829 & 86 & 60 & TKr \\
2003 June 11 & 52802.0140 & 52802.1140 & 92 & 20 & SY \\
 & 52802.1569 & 52802.3072 & 61 & 60 & TKr \\
 & 52802.3782 & 52802.5009 & 126 & 45 & JP \\
 & 52802.3848 & 52802.5333 & 203 & 50 & TV \\
2003 June 12 & 52802.7268 & 52802.8500 & 75 & 120 & MA \\
 & 52803.1551 & 52803.3364 & 122 & 60 & TKr \\
2003 June 13 & 52803.9766 & 52804.0338 & 62 & 30 & NH \\
 & 52804.1516 & 52804.3296 & 123 & 60 & TKr \\
 & 52804.3974 & 52804.5200 & 167 & 50 & TV \\
2003 June 14 & 52805.1537 & 52805.2685 & 78 & 60 & TKr \\
2003 June 15 & 52805.5862 & 52805.7097 & 105 & 90 & DS \\
 & 52805.9736 & 52806.0933 & 246 & 20 & SY \\
 & 52806.1573 & 52806.3333 & 107 & 120 & TKr \\
2003 June 16 & 52806.5884 & 52806.7076 & 92 & 90 & DS\\
 & 52807.1613 & 52807.3274 & 58 & 120 & TKr \\
2003 June 17 & 52808.1572 & 52808.3105 & 54 & 240 & TKr \\
2003 June 18 & 52808.6183 & 52808.7090 & 71 & 90 & DS \\
2003 June 19 & 52810.6007 & 52810.6362 & 31 & 90 & DS \\
\hline 
\multicolumn{6}{l}{$^*$ HJD-2400000} \\
\multicolumn{6}{l}{$^\dagger$ Number of frames} \\
\multicolumn{6}{l}{$^\ddagger$ Exposure times} \\
\multicolumn{6}{l}{Instrument TO:25cm+Apogee AP-7(Saitama, Japan),} \\
\multicolumn{6}{l}{GU:25cm+AP7p (Gunma, Japan),} \\
\multicolumn{6}{l}{ND:60cm+PixCellentS/T 00-3194(SITe 003AB)(Hida, Japan)} \\
\multicolumn{6}{l}{KYO:30cm+ST-7E(Kyoto, Japan)} \\
\multicolumn{6}{l}{LC:73cm+GenesisG16+90KAF1602E(California, USA)} \\ 
\multicolumn{6}{l}{NH:25cm+CV-04(Mie, Japan)} \\
\multicolumn{6}{l}{TA:21cm+ST-7(Okayama, Japan)} \\
\multicolumn{6}{l}{SY:28cm+ST-9E(Nayoro, Japan)} \\
\multicolumn{6}{l}{TV:35cm+ST-7 (Landen, Belgium)} \\
\multicolumn{6}{l}{DS:36cm+ST-10XME(Indiana, USA)} \\
\multicolumn{6}{l}{BY:45cm+PRIMEFOCUS (Arizona, USA)} \\
\multicolumn{6}{l}{JP:20cm+ST-6B (Erftstadt, Germany)} \\
\multicolumn{6}{l}{MA:12.5cm+ST-7E (Alberta, Canada)} \\
\multicolumn{6}{l}{TKr:28cm+ST-7E (Tashkent, Uzbekistan)} \\
\end{longtable}

\begin{longtable}{cccc}
\caption{Recorded outbursts of GO Com}
\hline\hline
day & JD & type$^*$ & references$^\dagger$ \\ 
\hline
\endhead
\hline
\endfoot
Jul. 1973 & 2441864 &? & 1 \\
May. 1989 & 2447676 & S? & 2 \\
Apr. 1990 & 2448011 & N? & 3 \\
Jul. 1995 & 2449914 & S & 4 \\
Mar. 1996 & 2450168 & N & 2 \\
Feb. 1997 & 2450486 & N & 2 \\
Apr. 1998 & 2450558 & N & 2 \\
Jun. 2003 & 2452794 & S & 5 \\
\hline
\multicolumn{3}{l}{$^*$ S: superoutburst, N: normal outburst } \\
\multicolumn{3}{l}{$^\dagger$ 1. \cite{kow77gocomiauc} 2. VSNET archives
 3. \cite{kat90gocom} 4. \cite{kat95gocom} 5. this work} \\
\end{longtable}

\begin{table}[htb]
\caption{Timing of superhumps maxima during the superoutburst}
\begin{center}
\begin{tabular}{rcrrcccc}
\hline\hline
HJD-2400000 & $E$ & $O-C$ \\ 
\hline
52795.1571 & 0 & 0.0055 \\
52795.2184 & 1 & 0.0038 \\
52795.2807 & 2 & 0.0030 \\
52795.7230 & 9 & 0.0039 \\
52795.7843 & 10 & 0.0021 \\
52796.0395 & 14 & 0.0051 \\
52796.0995 & 15 & 0.0020 \\
52796.1608 & 16 & 0.0003 \\
52796.4147 & 20 & 0.0019 \\
52796.4776 & 21 & 0.0018 \\
52796.5402 & 22 & 0.0013 \\
52796.7282 & 25 & 0.0001 \\
52796.7889 & 26 & -0.0022 \\
52796.8531 & 27 & -0.0011 \\
52797.0425 & 30 & -0.0009 \\
52797.4186 & 36 & -0.0031 \\
52797.4800 & 37 & -0.0048 \\
52797.8025 & 42 & 0.0024 \\
52798.1744 & 48 & -0.0041 \\
52798.2365 & 49 & -0.0050 \\
52798.2993 & 50 & -0.0053 \\
52798.4276 & 52 & -0.0031 \\
52798.4907 & 53 & -0.0031 \\
52798.6747 & 56 & -0.0082 \\
52798.9929 & 61 & -0.0053 \\
52799.1167 & 63 & -0.0076 \\
52799.1794 & 64 & -0.0080 \\
52799.2437 & 65 & -0.0068 \\
52799.3146 & 66 & 0.0011 \\
52799.7486 & 73 & -0.0063 \\
52800.0104 & 77 & 0.0032 \\
52800.0753 & 78 & 0.0051 \\
52800.1928 & 80 & -0.0036 \\
52800.3299 & 82 & 0.0074 \\
52801.2208 & 96 & 0.0154 \\
52802.4025 & 115 & -0.0010 \\
52802.4651 & 116 & -0.0014 \\
\hline
\end{tabular}
\end{center}
\end{table}

\begin{table}
\caption{Fundamental characteristics of GO Com and LL And}
\begin{tabular}{ccccccc}
\hline\hline
 & supercycle$^\dagger$ & plateau$^\dagger$ & amplitude &
 $P_{orb}$$^\dagger$ & $P_{sh}$$^\dagger$ & references$^\ddagger$ \\
\hline
GO Com & $\sim$ 2800 & 8 & 4.8 &  ? & 0.06359 & 1 \\
LL And & $\sim$ 5000 & 9 & $\sim$ 5 & 0.05505 & 0.05697 & 2 \\
\hline 
\multicolumn{3}{l}{$^\dagger$ unit in day} \\
\multicolumn{3}{l}{$^\ddagger$ 1. This work 2. \citet{how94lland},
 \citet{kat04lland}} \\
\end{tabular}
\end{table}


\begin{thebibliography}{}

\bibitem[Baba et~al.(2000)]{bab00v1028cyg}
  Baba, H., Kato, T., Nogami, D., Hirata, R., Matsumoto, K., \& Sadakane, K.\
  2000, \pasj, 52, 429

\bibitem[Bailey(1979)]{bai79wzsge}
  Bailey, J.\ 1979, \mnras, 189, 41P

\bibitem[Belyavskij(1933)]{bel33gocom}
  Belyavskij, A.~I.\ 1933, Perem. Zvezdy, 4, 234

\bibitem[Downes, Margon(1981)]{dow81wzsge}
  Downes, R.~A., \& Margon, B.\ 1981, \mnras, 197, 35P

\bibitem[Hellier(2001)]{hel01book}
  Hellier, C.\ 2001, Cataclysmic Variable Stars: how and why they vary
  (Berlin: Springer-Verlag)

\bibitem[Howell, Hurst(1994)]{how94lland}
  Howell, S.~B., \& Hurst, G.~M.\ 1994, Inf. Bull. Variable Stars, 4043

\bibitem[Howell et~al.(1995a)]{how95TOAD}
  Howell, S.~B., Szkody, P., \& Cannizzo, J.~K.\ 1995a, \apj, 439, 337

\bibitem[Howell et~al.(1990)]{how90faintCV3}
  Howell, S.~B., Szkody, P., Kreidl, T.~J., Mason, K.~O., \& Puchnarewicz,
  E.~M.\ 1990, \pasp, 102, 758

\bibitem[Howell et~al.(1995b)]{how95swumabcumatvcrv}
  Howell, S.~B., Szkody, P., Sonneborn, G., Fried, R., Mattei, J., Oliversen,
  R.~J., Ingram, D., \& Hurst, G.~M.\ 1995b, \apj, 453, 454

\bibitem[Hu et~al.(1998)]{hu98j2329}
  Hu, J, Wei, J., Xu, D., Cao, L., \& Dong, X.\ 1998, Ann. Shangahi Obs., Acad.
  Sin., 19, 235

\bibitem[Ishioka et~al.(2003)]{ish03hvvir}
  Ishioka, R., {et~al.}\ 2003, \pasj, 55, 683

\bibitem[Jurcevic et~al.(1994)]{jur94bzuma}
  Jurcevic, J.~S., Honeycutt, R.~K., Schlegel, E.~M., \& Webbink, R.~F.\ 1994,
  \pasp, 106, 481

\bibitem[Kaluzny(1986)]{kal86bzumaiauc}
  Kaluzny, J.\ 1986, \iaucirc, 4287

\bibitem[Kato(1997)]{kat97tleo}
  Kato, T.\ 1997, \pasj, 49, 583

\bibitem[Kato(1999)]{kat99bzuma}
  Kato, T.\ 1999, Inf. Bull. Variable Stars, 4768

\bibitem[Kato(2004)]{kat04lland}
  Kato, T.\ 2004, \pasj, 56, S135

\bibitem[Kato, Hirata(1990)]{kat90gocom}
  Kato, T., \& Hirata, R.\ 1990, Inf. Bull. Variable Stars, 3489

\bibitem[Kato et~al.(2001a)]{kat01wxcet}
  Kato, T., Matsumoto, K., Nogami, D., Morikawa, K., \& Kiyota, S.\ 2001a,
  \pasj, 53, 893

\bibitem[Kato et~al.(1995)]{kat95gocom}
  Kato, T., Nogami, D., \& Baba, H.\ 1995, Inf. Bull. Variable Stars, 4228

\bibitem[Kato et~al.(2003a)]{kat03hodel}
  Kato, T., Nogami, D., Moilanen, M., \& Yamaoka, H.\ 2003a, \pasj, 55, 989

\bibitem[Kato et~al.(2003b)]{kat03v877arakktelpucma}
  Kato, T., {et~al.}\ 2003b, \mnras, 339, 861

\bibitem[Kato et~al.(2001b)]{kat01hvvir}
  Kato, T., Sekine, Y., \& Hirata, R.\ 2001b, \pasj, 53, 1191

\bibitem[Kato, Starkey(2002)]{kat02v592cas}
  Kato, T., \& Starkey, D.~R.\ 2002, Inf. Bull. Variable Stars, 5358

\bibitem[Kato et~al.(2004)]{VSNET}
  Kato, T., Uemura, M., Ishioka, R., Nogami, D., Kunjaya, C., Baba, H., \&
  Yamaoka, H.\ 2004, \pasj, 56, S1

\bibitem[Kowal(1973)]{kow77gocomiauc}
  Kowal, C.~T.\ 1973, \iaucirc, 2562

\bibitem[Lasota et~al.(1995)]{las95wzsge}
  Lasota, J.~P., Hameury, J.~M., \& Hur\'{e}, J.~M.\ 1995, \aap, 302, L29

\bibitem[Marino, Walker(1979)]{mar79superoutburst}
  Marino, B.~F., \& Walker, W. S.~G.\ 1979, in IAU Colloq. 46, Changing Trends
  in Variable Star Research, ed. F.~M. Bateson, J. Smak, \& J.~H. Urch (Univ.
  of Waikato, Hamilton), N. Z., 29

\bibitem[Mineshige et~al.(1998)]{min98wzsge}
  Mineshige, S., Liu, B., Meyer, F., \& Meyer-Hofmeister, E.\ 1998, \pasj, 50,
  L5

\bibitem[Moreau, Reboul(1995)]{mor95quasar}
  Moreau, O., \& Reboul, H.\ 1995, \aaps, 111, 169

\bibitem[Neustroev et~al.(2002)]{neu02bzumaproc}
  Neustroev, V.~V., Medvedev, A., Turbin, S., \& Borisov, N.~V.\ 2002, in
  ASP Conf. Ser. 261, The Physics of Cataclysmic Variables and Related Objects, ed.
  B.~T. G\"{a}nsicke, K. Beuermann, \& K. Reinsch (San Francisco: ASP), 515

\bibitem[Nogami et~al.(1998)]{nog98swuma}
  Nogami, D., Baba, H., Kato, T., \& Nov\'{a}k, R.\ 1998, \pasj, 50, 297

\bibitem[Nogami et~al.(1997)]{nog97sxlmi}
  Nogami, D., Masuda, S., \& Kato, T.\ 1997, \pasp, 109, 1114

\bibitem[Nogami et~al.(2004)]{nog04vwcrb}
  Nogami, D., Uemura, M., Ishioka, R., Kato, T., \& Pietz, J.\ 2004, \pasj, 56,
  S155

\bibitem[O'Donoghue et~al.(1991)]{odo91wzsge}
  O'Donoghue, D., Chen, A., Marang, F., Mittaz, J. P.~D., Winkler, H., \&
  Warner, B.\ 1991, \mnras, 250, 363

\bibitem[Olech(2003)]{ole03v1141aql}
  Olech, A.\ 2003, Acta Astron., 53, 85

\bibitem[Olech et~al.(2003)]{ole03ksuma}
  Olech, A., Schwarzenberg-Czerny, A., K\c{e}dzierski, P., Z{\l}oczewski, K.,
  Mularczyk, K., \& Wi\'{s}niewski, M.\ 2003, Acta Astron., 53, 175

\bibitem[Osaki(1985)]{osa85SHexcess}
  Osaki, Y.\ 1985, \aap, 144, 369

\bibitem[Osaki(1995)]{osa95wzsge}
  Osaki, Y.\ 1995, \pasj, 47, 47

\bibitem[Osaki(1996)]{osa96review}
  Osaki, Y.\ 1996, \pasp, 108, 39

\bibitem[Osaki, Meyer(2002)]{osa02wzsgehump}
  Osaki, Y., \& Meyer, F.\ 2002, \aap, 383, 574

\bibitem[Osaki, Meyer(2003)]{osa03DNoutburst}
  Osaki, Y., \& Meyer, F.\ 2003, \aap, 401, 325

\bibitem[Patterson(2001)]{pat01SH}
  Patterson, J.\ 2001, \pasp, 113, 736

\bibitem[Patterson et~al.(1993)]{pat93vyaqr}
  Patterson, J., Bond, H.~E., Grauer, A.~D., Shafter, A.~W., \& Mattei, J.~A.\
  1993, \pasp, 105, 69

\bibitem[Patterson et~al.(2002)]{pat02wzsge}
  Patterson, J., {et~al.}\ 2002, \pasp, 114, 721

\bibitem[Ringwald, Thorstensen(1990)]{rin90bzuma}
  Ringwald, F.~A., \& Thorstensen, J.~R.\ 1990, \baas, 22, 1291

\bibitem[Semeniuk et~al.(1997)]{sem97swuma}
  Semeniuk, I., Olech, A., Kwast, T., \& Nalezyty, M.\ 1997, Acta Astron., 47, 201

\bibitem[Stellingwerf(1978)]{ste78pdm}
  Stellingwerf, R.~F.\ 1978, \apj, 224, 953

\bibitem[Uemura et~al.(2002a)]{uem02j2329}
  Uemura, M., {et~al.}\ 2002a, \pasj, 54, 599

\bibitem[Uemura et~al.(2002b)]{uem02j2329letter}
  Uemura, M., {et~al.}\ 2002b, \pasj, 54, L15

\bibitem[Usher(1981)]{ush81gocom}
  Usher, P.~D.\ 1981, \apjs, 46, 117

\bibitem[Verbunt et~al.(1997)]{ver97ROSAT}
  Verbunt, F., Bunk, W.~H., Ritter, H., \& Pfeffermann, E.\ 1997, \aap, 327,
  602

\bibitem[Vogt, Bateson(1982)]{vog82atlas}
  Vogt, N., \& Bateson, F.~M.\ 1982, \aaps, 48, 383

\bibitem[Warner(1985)]{war85suuma}
  Warner, B.\ 1985, in Interacting Binaries, ed. P.~P. Eggleton, \& J.~E.
  Pringle (Dordrecht: D. Reidel Publishing Company), ~367

\bibitem[Warner(1995)]{war95book}
  Warner, B.\ 1995, Cataclysmic Variable Stars (Cambridge: Cambridge University Press)

\bibitem[Wei et~al.(2001)]{wei01j2329}
  Wei, J., Jiang, X., Xu, D., Zhou, A., \& Hu, J.\ 2001,
  Chin.J.Astron.Astrophys, 1, 483

\bibitem[Wenzel(1982)]{wen82bzuma}
  Wenzel, W.\ 1982, Inf. Bull. Variable Stars, 2256

\end{thebibliography}
\end{document}